\documentstyle{mn2e}
\begin{document}
\def\simlt{\mathrel{\rlap{\lower 3pt\hbox{$\sim$}}
        \raise 2.0pt\hbox{$<$}}}
\def\simgt{\mathrel{\rlap{\lower 3pt\hbox{$\sim$}}
        \raise 2.0pt\hbox{$>$}}}
\def\bj{b_{\rm\scriptscriptstyle J}}
\def\rt{r_{\rm\scriptscriptstyle T}}
\def\rp{r_{\rm\scriptscriptstyle P}}
\renewcommand{\labelenumi}{(\arabic{enumi})}

\title[The 2dF Galaxy Redshift Survey: Clustering properties of radio galaxies]
{The 2dF Galaxy Redshift Survey: Clustering properties of radio galaxies}
\author[Manuela Magliocchetti et al. (the 2dFGRS Team)]
{\parbox[t]\textwidth{
Manuela Magliocchetti$^{1}$,
Steve J.\ Maddox${^2}$, 
Ed Hawkins${^2}$,
John A.\ Peacock$^{3}$,
Joss Bland-Hawthorn$^4$,
Terry Bridges$^4$,
Russell Cannon$^4$,
Shaun Cole$^5$,
Matthew Colless$^6$,
Chris Collins$^7$,
Warrick Couch$^8$,
Gavin Dalton$^{10}$,
Roberto de Propris$^8$,
Simon P.\ Driver$^9$,
George Efstathiou$^{11}$,
Richard S.\ Ellis$^{12}$,
Carlos S.\ Frenk$^5$,
Karl Glazebrook$^{13}$,
Carole A.\ Jackson${^6}$,
Bryn Jones${^3}$, 
Ofer Lahav$^{11}$,
Ian Lewis$^5$,
Stuart Lumsden$^{14}$,
Peder Norberg$^{15}$,
Bruce A.\ Peterson$^4$,
Will Sutherland$^{3}$,
Keith Taylor$^4$ (the 2dFGRS Team)}
\vspace*{6pt} \\
{\tt $^1$SISSA, Via Beirut 4, 34100, Trieste, Italy} \\
{\tt $^2$School of Physics and Astronomy, University of Nottingham,
Nottingham NG7 2RD, UK}\\
{\tt $^{3}$Institute for Astronomy, University of Edinburgh, Royal 
       Observatory, Blackford Hill, Edinburgh EH9 3HJ, UK}\\
{\tt $^4$Anglo-Australian Observatory, P.O.\ Box 296, Epping, NSW 2121,
    Australia}\\
{\tt $^5$Department of Physics, University of Durham, South Road,
    Durham DH1 3LE, UK} \\
{\tt $^6$Research School of Astronomy and Astrophysics,
 The Australian National University, Canberra, ACT 2611, Australia}\\
{\tt $^7$Astrophysics Research Institute, Liverpool John Moores University,
    Twelve Quays House, Birkenhead, L14 1LD, UK} \\
{\tt $^8$Department of Astrophysics, University of New South Wales, Sydney,
    NSW 2052, Australia} \\
{\tt $^9$School of Physics and Astronomy, University of St Andrews,
    North Haugh, St Andrews, Fife, KY6 9SS, UK} \\
{\tt $^{10}$Department of Physics, University of Oxford, Keble Road,
    Oxford OX1 3RH, UK} \\
{\tt $^{11}$Institute of Astronomy, University of Cambridge, Madingley Road,
    Cambridge CB3 0HA, UK} \\
{\tt $^{12}$Department of Astronomy, California Institute of Technology,
    Pasadena, CA 91125, USA} \\
{\tt $^{13}$Department of Physics \& Astronomy, Johns Hopkins University,
       Baltimore, MD 21218-2686, USA} \\
{\tt $^{14}$Department of Physics, University of Leeds, Woodhouse Lane,
       Leeds, LS2 9JT, UK} \\
{\tt $^{15}$ETHZ Institut f\"ur Astronomie, HPF G3.1, ETH H\"onggerberg, CH-8093
       Z\"urich, Switzerland} \\}
\maketitle

\vspace{7cm}
\begin{abstract}
The clustering properties of local, $S_{1.4 \rm GHz}\ge 1$~mJy, radio sources 
are investigated for a sample of 820 objects drawn from the joint use of the 
FIRST and 2dF Galaxy Redshift surveys. To this aim, we present 271 new 
$\bj \le 19.45$ spectroscopic counterparts of FIRST radio sources to be added 
to those already introduced in Magliocchetti et al. (2002).
The two-point correlation function for the local radio population is found 
to be entirely consistent with estimates obtained for the whole sample 
of 2dFGRS galaxies. From measurements of the redshift-space correlation function 
$\xi(s)$ we derive a redshift-space clustering length $s_0=10.7^{+0.8}_
{-0.7}$~Mpc,
while from the projected correlation function $\Xi(\rt)$ we estimate 
the parameters of the real-space correlation function $\xi(r)=(r/r_0)^
{-\gamma}$, $r_0=6.7^{+0.9}_{-1.1}$~Mpc and $\gamma=1.6\pm 0.1$,
where $h=0.7$ is assumed. Different results are 
instead obtained if we only consider sources that present signatures 
of AGN activity in their spectra. These objects are shown to be very strongly 
correlated, with $r_0=10.9^{+1.0}_{-1.2}$~Mpc and $\gamma=2\pm 0.1$, 
a steeper slope than has been claimed in other recent works.
No difference is found in the clustering 
properties of radio-AGNs of different radio luminosity. 
Comparisons with models for $\xi(r)$ show that AGN-fuelled sources reside in 
dark matter halos more massive than $\rm{\sim 10^{13.4} M_{\sun}}$,
higher the corresponding figure for radio-quiet QSOs. This value can be converted 
into a minimum black hole mass associated with
radio-loud, AGN-fuelled objects of $\rm M_{BH}^{min}\sim 10^9 M_{\sun}$.
The above results then suggest -- at least for relatively faint radio objects 
-- the existence of a threshold
black hole mass associated with the onset of 
significant radio activity such as that of radio-loud AGNs; however, 
once the activity is triggered, there appears to be no evidence for a 
connection between black hole mass and level of radio output.    
\end{abstract}


\section{Introduction}

The last twenty years have shown radio sources to be extremely good
probes of cosmological large-scale structure up to significant ($z\sim
4$) redshifts.  Clustering in radio catalogues was detected in early
wide-area studies: Seldner \& Peebles (1981) and Shaver \& Pierre
(1989) reported the detection of slight clustering of nearby radio
sources, while Kooiman, Burns \& Klypin (1995) and Loan, Wall \& Lahav
(1997) measured strong anisotropy in the distribution of bright radio
objects from the 4.85~GHz Green Bank and Parkes-MIT-NRAO surveys.
However, it was only few years ago that the latest generation of radio
surveys such as FIRST (Faint Images of the Radio Sky at 20 cm; Becker,
White \& Helfand 1995), WENSS (Rengelink et al. 1998) and NVSS (NRAO
VLA Sky Survey; Condon et al. 1998) included enough objects to allow
for high-precision clustering measurements (Cress et al. 1996;
Rengelink et al. 1998; Magliocchetti et al. 1998; Blake \& Wall 2003;
Overzier et al. 2003). All these recent analyses reveal the tendency
for radio objects to be more strongly clustered than
optically-selected galaxies.

Assessing the real clustering signal is nevertheless not an easy task 
when it comes to radio sources, since the lack of known redshifts for the 
overwhelming majority of these objects forces one to deal with projected 
quantities such as the angular two-point correlation function $w(\theta)$; 
unfortunately, the relation between angular and spatial measurements is 
dependent on the radio source redshift distribution $N(z)$ -- which 
is highly uncertain at the mJy flux levels probed by these new 
surveys (see e.g. Dunlop \& Peacock 1990; Magliocchetti et al. 
1999) -- and on the unknown redshift evolution of the radio clustering signal. 
This leads to estimates for the comoving correlation length $r_0$ 
(here we are assuming the spatial two-point correlation function to be 
described by the power-law $\xi(r)=(r/r_0)^{- \gamma}$) which may span the 
relatively broad range $\sim 7-15$~Mpc even for angular measurements in good 
agreement with each other (Magliocchetti et al. 1998; Blake \& Wall 2003; 
Overzier et al. 2003). 

Since the natural solution to the above problem (redshift acquisition for 
all the objects included in a chosen catalogue) is out of reach with 
current instruments, a good starting point towards the understanding of the 
clustering properties of radio sources is the analysis of homogeneous 
subsamples of objects, bright enough in the visible band to allow for optical 
and spectroscopic follow-ups. 

A first attempt in this direction was performed by Peacock \& Nicholson (1991) 
who measured the redshift-space correlation function for a sample of 310 
radio galaxies with $z\simlt 0.1$ and radio fluxes $S>0.5$~Jy at 1.4~GHz.
These authors indeed found radio sources in their catalogue to be strongly 
clustered, with a redshift-space correlation length $s_0\simeq 11 h^{-1}$~Mpc. 

The present paper analyses the clustering properties of $\sim 820$, 
$z\simlt 0.3$, $S_{1.4 \rm GHz}\ge 1$~mJy radio galaxies drawn from the joint 
use of the FIRST and 2dF Galaxy Redshift surveys as illustrated in 
Magliocchetti et al. (2002). By doing this, we not only extend the 
Peacock \& Nicholson (1991) measurements to a statistically more significant 
sample involving less local objects, but we also probe much lower flux densities 
where the population contains radio-emitting sources that differ from typical AGNs 
(such as galaxies undergoing intense star formation).

In addition to the above analysis, we will also estimate 
the two-point correlation function (both in redshift space and real space) for the 
homogeneous sample of radio AGNs (sources that present signatures of AGN 
activity in their optical spectra). This will enable us to use up-to-date 
models, which connect galaxy formation properties with clustering behaviour 
in order to investigate the nature of the dark matter halos in which these 
sources reside. By means of the same approach we can also obtain precious 
information on the nature of the black hole associated with a radio-active AGN 
to be compared with conclusions from other works. 

For instance, Auriemma et al. (1977) found that radio emission is more common in 
brighter/more massive ellipticals. This result was recently confirmed e.g. by findings of Falomo 
et al. (2003), while authors such as Laor (2000), Lacy et al. (2001), McLure \& Dunlop (2002), 
McLure \& Jarvis (2002) and Dunlop et al. (2003) have investigated the connection between 
black hole mass and radio luminosity, coming to conclusions in disagreement with each other.

The layout of the paper is as follows. Section 2 introduces 271 new 
spectroscopic counterparts for local, $S_{\rm 1.4~GHz}\ge 1$~mJy, 
radio sources, derived from the joint use of the FIRST and 2dF Galaxy 
Redshift (2dFGRS) surveys. 
Section 3.1 describes some of the properties of the total FIRST-2dFGRS sample, 
which puts together objects presented in Section 2 with those introduced in 
Magliocchetti et al. (2002), while Sections 3.2 and 3.3 are respectively 
devoted to the analysis of the redshift-space correlation function $\xi(s)$ 
and of the projected correlation function ${\Xi}(\rt)$ both for the 
whole radio-2dFGRS sample and only for the 536 radio-AGNs. Section 4 compares the 
available data with models in order to derive the minimum dark matter mass of 
a halo able to host an AGN-fuelled radio source and gives some constraints on 
the minimum black hole mass of a radio-loud AGN. Finally, Section 5 summarizes 
our conclusions.

Unless stated otherwise, throughout this work we will assume $\Omega_0=0.3$, 
$\Lambda=0.7$, $h=0.7$ and
$\sigma_8=0.8$ (with $\sigma_8$ the rms density fluctuation within a sphere
with a radius of $8 h^{-1}$ Mpc), as the latest results
from the joint analysis of CMB and 2dFGRS data seem to indicate (see e.g. Lahav
et al., 2002; Spergel et al. 2003). 
To avoid confusion, we especially emphasise that distances will thus normally be
quoted in Mpc, assuming $h=0.7$. Units of $h^{-1}$~Mpc will occur only rarely. 
This assumption is also made in all figures
quoted for radio luminosities and black-hole masses.
All the correlation lengths are in comoving units.

\section{More Spectroscopic Counterparts for FIRST Radio Sources}

\begin{figure*}
\vspace{8cm}  
\includegraphics{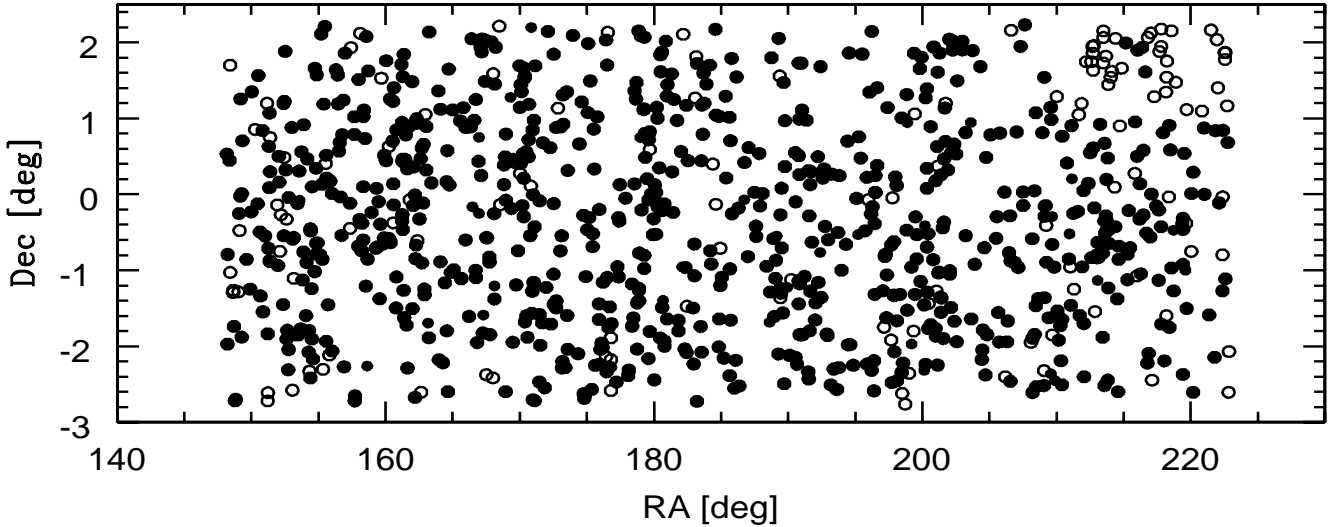}
\caption{Projected distribution of radio sources with optical counterparts
in the APM catalogue for $\bj\le 19.45$.
Filled circles identify objects with spectral identifications, while empty
ones are for those not included in the 2dFGRS.
\label{fig:fields}}
\end{figure*}

\begin{figure*}
\vspace{8cm}  
\includegraphics{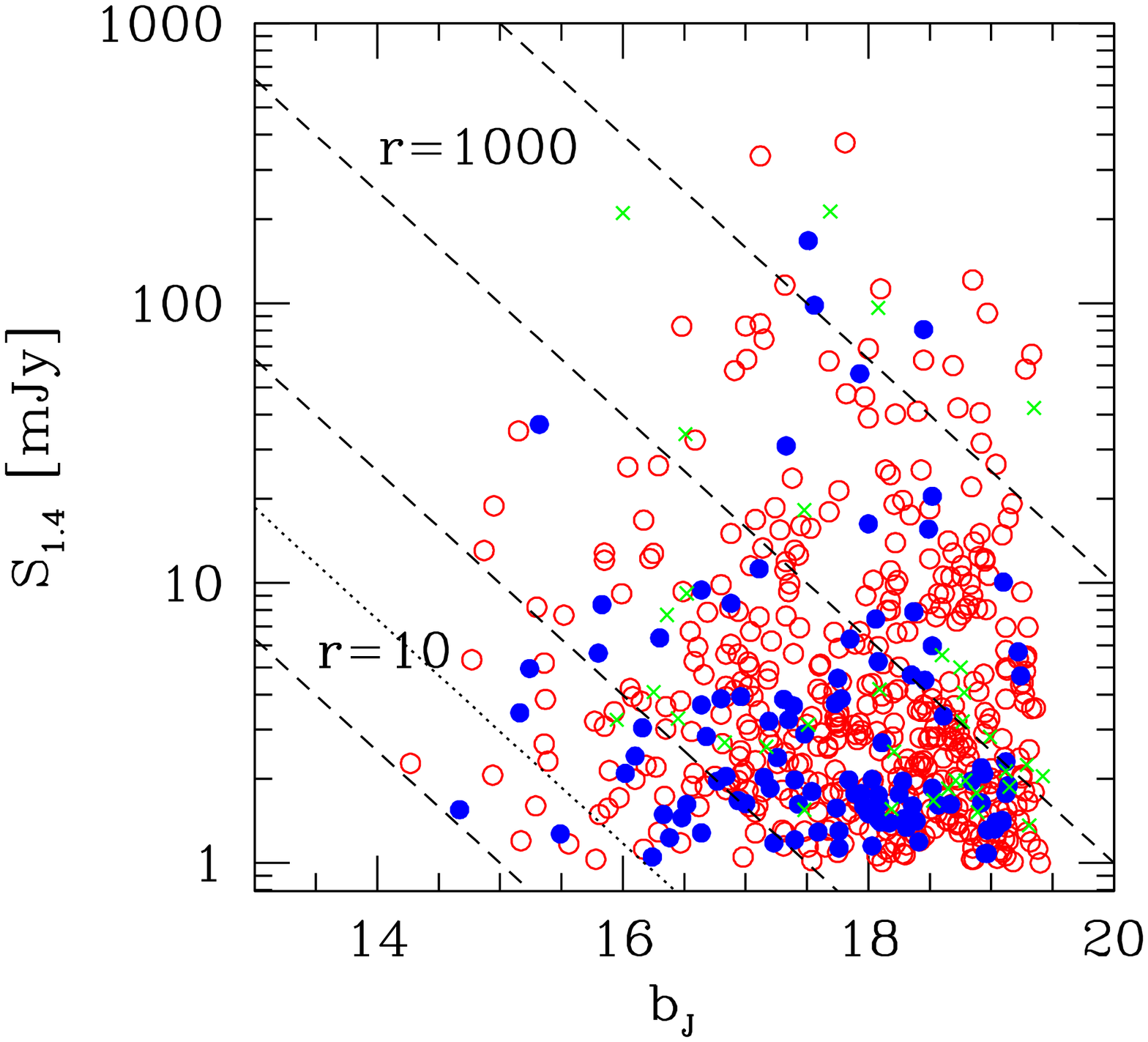}
\includegraphics{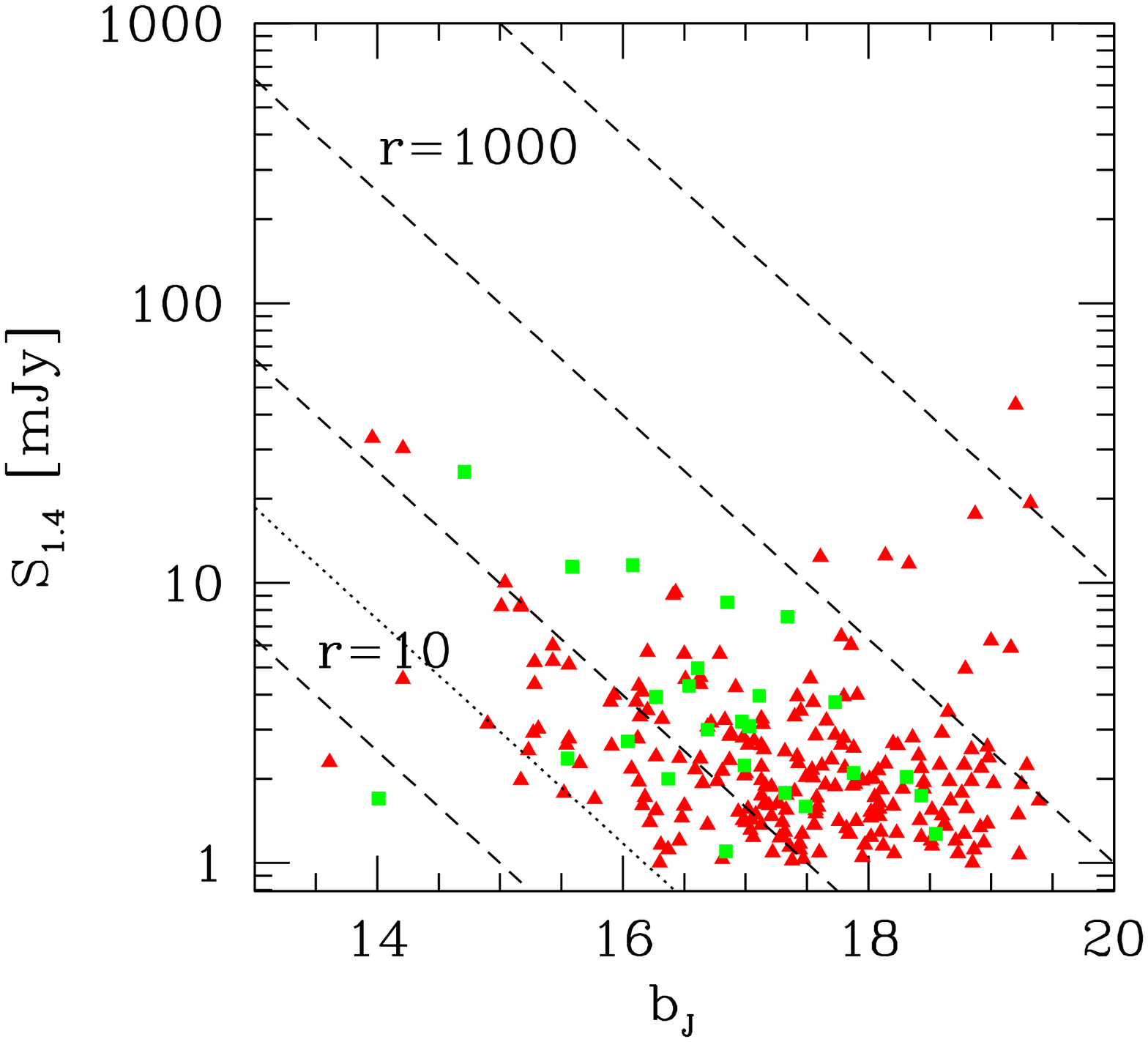}
\caption{$\bj$ magnitudes versus radio flux $S$ at 1.4~GHz for different 
classes of objects. The left-hand panel shows the case for 
early-type galaxies (represented by empty circles), E+AGN 
(filled circles) and Seyfert galaxies (crosses), while the right-hand panel 
is obtained for the populations of late-type galaxies (filled triangles) 
and starbursts (filled squares). Dashed lines correspond 
to constant values of the radio-to-optical ratio $r=10,100,10^3,10^4$; 
the dotted lines indicate the threshold of $r\sim 30$ for radio-loudness.
\label{fig:FB}}
\end{figure*}

The 2dFGRS (Colless et al. 2001, 2003)
is a large-scale survey selected in the photometric $\bj$ band from the 
APM catalogue (Maddox et al. 1990a, 1990b, 1996)  and its subsequent 
extensions (Maddox et al., in preparation). The final version includes 
221,414 unique, reliable (i.e. with quality flags $Q\ge 3$, Colless et al. 
2001, 2003) galaxy redshifts to an extinction-corrected limit for completeness of 
$\bj\simeq 19.45$ (with small variations in magnitude limit as a function of 
position over the sky) over an area of 2151 square degrees.
The survey geometry consists of two broad declination strips, a larger one
in the SGP covering the area $3^h 30^m\simlt {\rm RA}({\rm 2000})\simlt
21^h40^m$, $-37.5^\circ \simlt {\rm dec}({\rm 2000})\simlt -22.5^\circ$ and 
a smaller one set in the NGP with $9^h 50^m\simlt {\rm RA}({\rm 2000})\simlt 
14^h50^m$, $2.5^\circ \simlt{\rm dec}({\rm 2000}) \simlt -7.5^\circ$, plus 
100 random 2-degree fields spread uniformly over the 7000 square degrees 
of the APM catalogue in the southern Galactic hemisphere. The median
redshift of the galaxies is 0.11 and the great majority have
$z<0.3$.  

The completion of the 2dFGRS has allowed us to obtain 271 new spectroscopic 
counterparts for $S_{1.4 \rm GHz}\ge 1$~mJy radio objects to be added 
to the sample presented in Magliocchetti et al. (2002).
As in Magliocchetti et al. (2002), the parent radio dataset comes from 
matching together sources 
in the FIRST (Becker et al. 1995) and APM catalogues over the  
region of the sky between $9^h 48^m \simlt {\rm RA}({\rm 2000})
\simlt 14^h 32^m$ and $-2.77^\circ \simlt {\rm dec}({\rm 2000}) 
\simlt 2.25^\circ$, where these two surveys overlap. Magliocchetti \& Maddox 
(2002) find 4075 identifications -- corresponding to 16.7 per cent of the 
original radio sample -- in the APM catalogue for $\bj\le 22$ and 
for a matching radius of 2 arcsec, value chosen as the best compromise 
to maximize the number of real associations (expected to be $\sim 97\% $ of the FIRST-APM 
dataset), 
while limiting the number of random coincidences to a negligible $\sim 5$ per 
cent. 
971 objects (hereafter indicated as the 
photometric catalogue) in the parent dataset exhibit $\bj$ magnitudes 
brighter than 19.45, the limit of the 2dFGRS. 

We note that, even though the fine angular resolution of the FIRST survey implies that some 
of the flux coming from extended sources could be either resolved 
out or split into two or more components, leading to a systematic underestimate of 
the real flux densities of such sources, this effect has been 
partially corrected for by using the method developed by Magliocchetti et al. (1998) 
to combine multi-component objects. The above technique, together with the matching procedure, 
associates optical counterparts to the centroids of multi-component sources, thus minimizing the 
chances of missing an object with extended radio emission. Also, as already discussed in 
Magliocchetti et al. (2002), the correction to radio fluxes as measured by FIRST is 
$\sim 1~mJy$ in the case of sources brighter 
than 3~mJy, and about 30 per cent for sources with $1\,{\rm mJy}\le S_{1.4 {\rm GHz}}\simlt$
3~mJy (which mainly correspond to compact spiral galaxies and starbursts). These corrections 
have been shown not to affect any of the results obtained in both the previous (Magliocchetti 
et al. 2002) and current works.
 
The new spectroscopic counterparts were identified by searching in 
the 2dFGRS catalogue for objects not yet present in the Magliocchetti et al. 
(2002) sample with positions that differed by less than 2
arcsec (value of the diameter of each 2dF fibre) from positions of 
sources in the parent photometric dataset. The properties of these 271 objects 
are described in Table 2 at the end of the paper; for each of them the Table 
indicates:\\

\noindent
(1) Source number\\
(2) Right Ascension $\alpha$ (J2000) and (3) Declination $\delta$ (J2000) 
as measured in the 2dFGRS.\\
(4) Offset (expressed in arcsecs) between radio and optical 
    counterpart in the APM catalogue.\\
(5) Radio-flux density (in mJy units) at 1.4~GHz.\\
(6) Apparent $\bj$ and, when present, (7) R magnitudes of the optical
counterpart.\\
(8) Redshift.\\
(9) Spectral Classification.\\

Following Magliocchetti et al. (2002), classes for the optical
counterparts of radio sources (column 9 of Table 2) have been assigned
on the basis of their 2dF spectra.  Spectra have been compared with
known templates (see e.g. Kennicut 1992; McQuade et al. 1995) which
allowed galaxies to be divided into 6 broad categories:
\begin{enumerate}
\item {\it Early-type galaxies}, 150 sources where spectra were dominated by 
continua much
stronger than the intensity of any emission line. These objects can be
further divided into two sub-classes:\\
\hglue\parindent (i)  galaxies with absorption lines only.\\
\hglue\parindent (ii) galaxies with absorption lines + weak [O${\rm II}$] 
and H$\alpha$ emission lines denoting little star-formation activity.
\item {\it E+AGN-type galaxies}, 40 sources showing spectra typical of 
early-types
plus the presence of (narrow) emission lines such as [O${\rm II}$],
[O${\rm III}$], [N${\rm II}$] and [S${\rm II}$], which are strong if
compared to any Balmer line in emission and indicate the presence of large,
partially ionized transition regions as is the case in active galaxies.
\item {\it Late-type galaxies}, 51 sources where spectra show strong 
emission (mainly
Balmer) lines characteristic of star-formation activity, together with a
detectable continuum.
\item {\it Starburst galaxies}, 15 sources with optical spectra
characterized  by an almost negligible continuum with very strong
emission lines indicating the presence of intense star-formation activity.
\item {\it Seyfert 1 galaxies}, 1 source with spectrum showing strong, broad 
emission lines.
\item {\it Seyfert 2 galaxies}, 9 sources where the continuum is missing and 
spectra
only show strong narrow emission lines due to the presence of an active
galactic nucleus.
\end{enumerate}

\begin{figure}
\vspace{8cm}  
\includegraphics{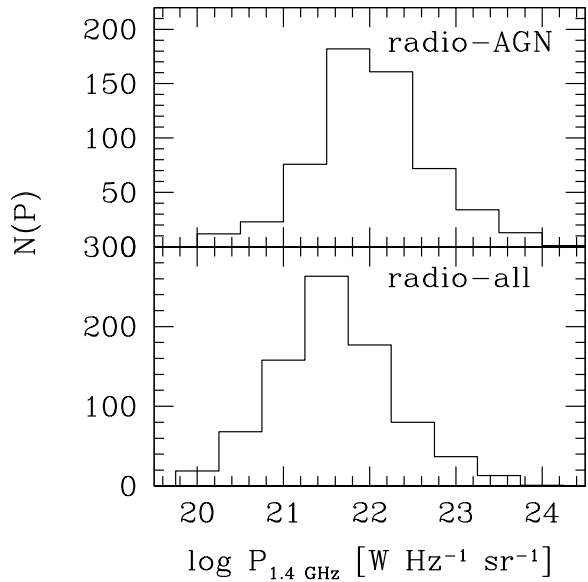}
\caption{Distribution of monochromatic radio power at 1.4~GHz for those
FIRST radio sources identified in the 2dFGRS survey. The top panel only 
includes 
objects that present signatures of AGN activity in their optical spectra, 
while the bottom panel is for the whole FIRST-2dFGRS sample. 
\label{fig:power_hist_type}}
\end{figure}

Distinctions between different classes of sources and in particular
between E+AGN, Seyfert 2 and Late-type galaxies have relied on the
diagnostic emission line ratios of Veilleux \& Osterbrock (1987),
Woltjer (1990) and Rola Terlevich \& Terlevich (1997). Note that a
definite classification was not possible for all the cases. This
simply reflects the fact that it is in general quite common to find
`composite' galaxies containing both an AGN and ongoing star formation
(see e.g. Hill et al. 2001). We also found two intermediate cases of
Seyfert 2-spectral-types with mildly broadened emission lines.

More details on the properties of the full FIRST-2dFGRS sample are given in 
Section 3.1.

\section{The Clustering Properties of Local Radio Sources}

\begin{figure*}
\vspace{8cm}  
\includegraphics{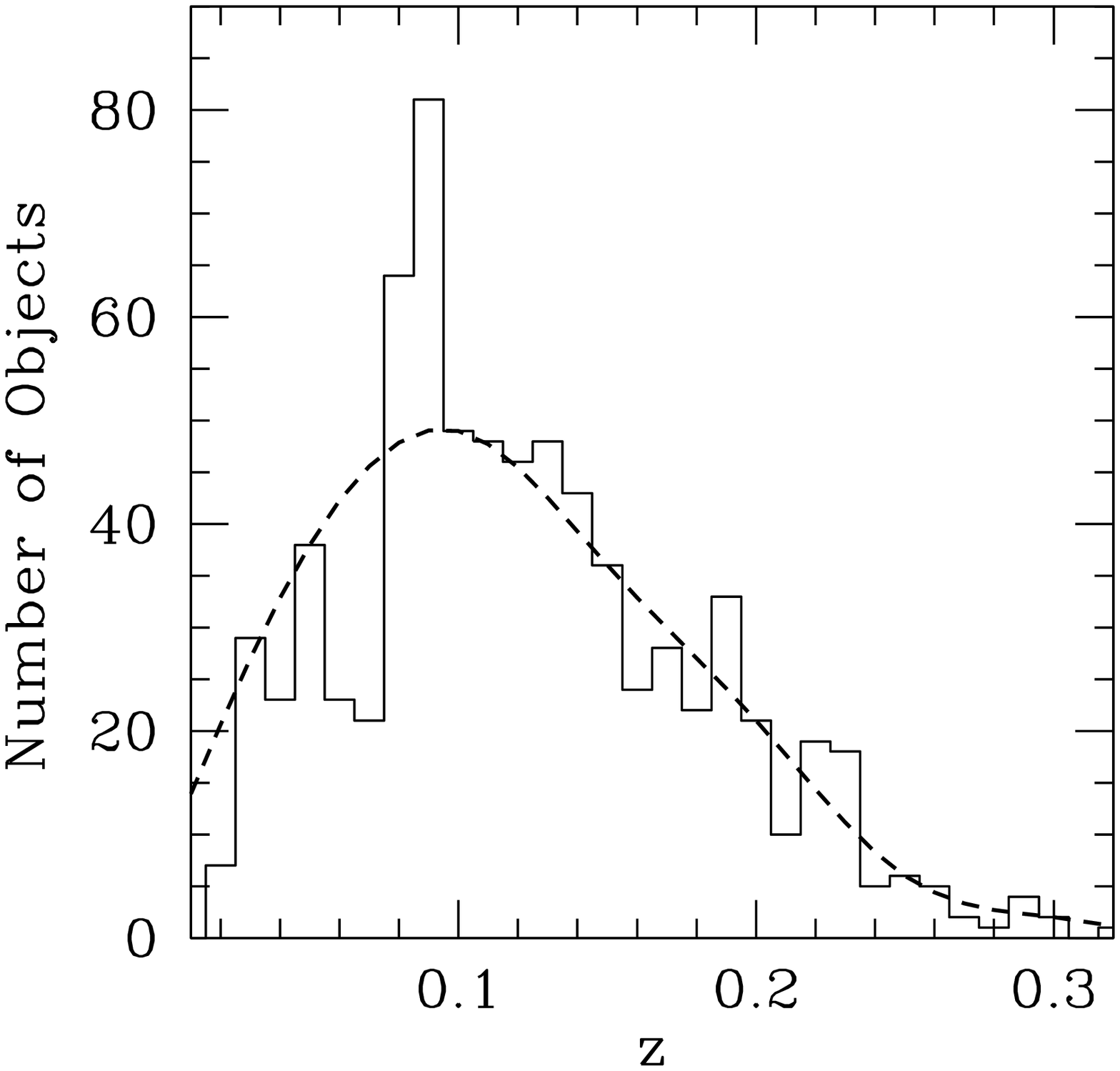}
\includegraphics{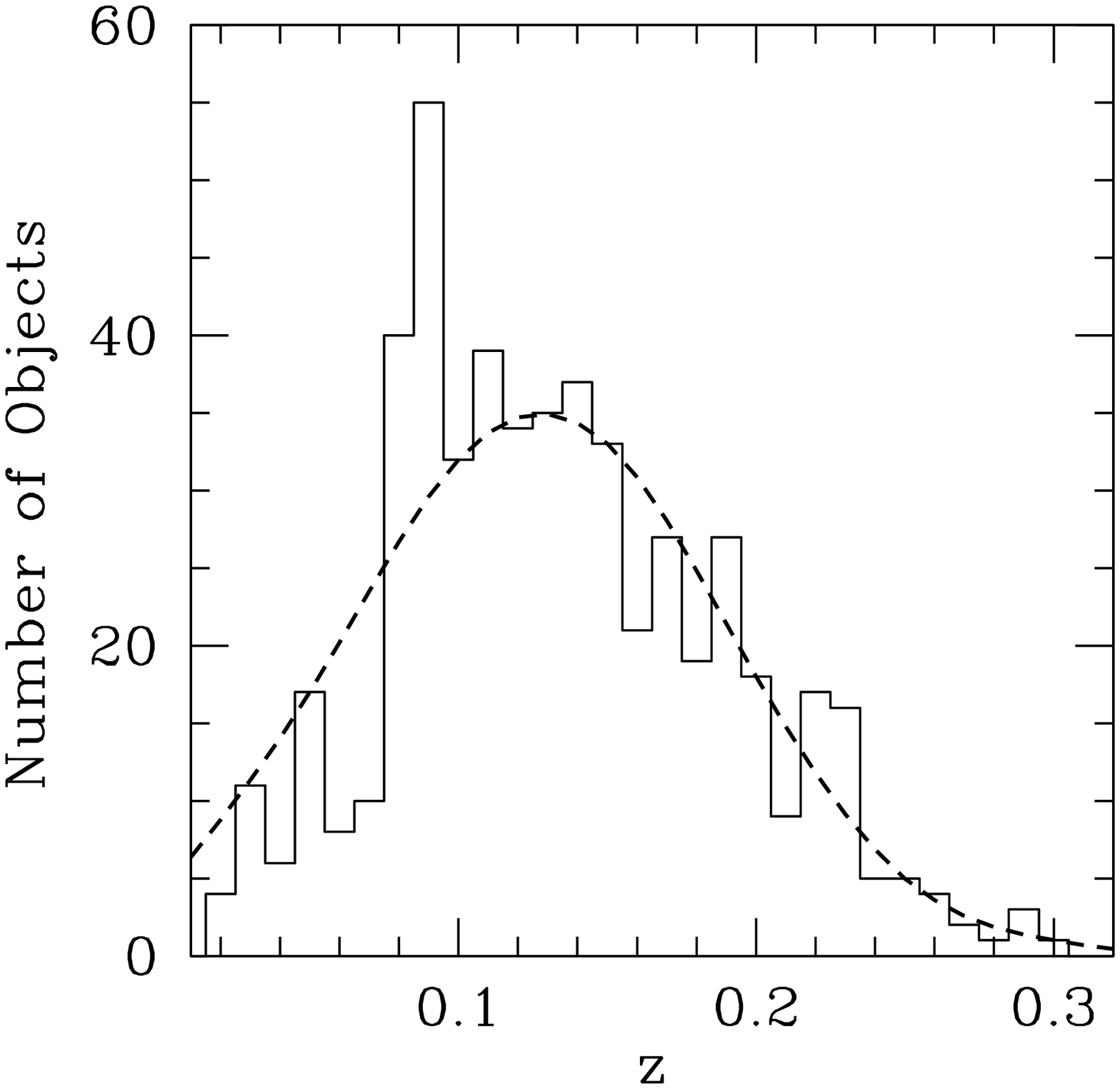}
\caption{Left-hand panel: redshift distribution $N(z)$ for the whole 
spectroscopic sample (solid line) and for the normalized random catalogues 
(dashed line). Right-hand panel: same as before but for the class of 
radio-AGN sources (see text for details).
\label{fig:N_z}}
\end{figure*}

\subsection{The Data Set}

The combination of the sample introduced in Magliocchetti et
al. (2002) and of those sources presented in Section 2 provides us
with a total of 828 FIRST radio objects with a spectroscopic
counterpart in the 2dFGRS.  This number is reduced to 820 if we only
consider spectra with quality flag Q$\ge$3 and discard the (four)
stars present in the dataset.  Their projected distribution onto the
sky is shown in Fig. \ref{fig:fields}, whereby empty circles
identify the 971 $\bj\le19.45$ and $S\ge 1$~mJy sources drawn from the
parent photometric catalogue (see Section 2), while filled dots
represent the 820 objects with available redshift estimates and
spectral classifications.

In order to investigate the radio properties of the sample, Figs
\ref{fig:FB} and \ref{fig:power_hist_type} respectively show $\bj$
magnitudes versus radio flux $S$ at 1.4~GHz and radio power
distribution. In more detail, the left-hand panel of Fig.
\ref{fig:FB} is devoted to the $\bj-S$ distribution of ``classical'',
AGN-powered sources (hereafter called radio-AGNs), while the
right-hand panel was obtained for the sub-class of star-forming
galaxies which owe their radio emission to processes different from
accretion onto a central black hole. Empty circles then are for
early-type galaxies, filled dots for E+AGN, crosses for Seyfert
galaxies (regardless of whether type 1 or 2), while filled triangles
are for late-type galaxies and filled squares for starbursts, where
the classification follows the one introduced in Section 2. The dashed
lines in both panels indicate the loci of constant radio-to-optical
ratios $r=S\times 10^{(\bj-12.5)/2.5}$, and the dotted lines
illustrate the threshold of $r\sim 30$ for an object to be considered
as radio-loud. It is clear from the plots that these two classes of
objects tend to occupy different regions of the $\bj-S$ plane,
star-forming sources being in general brighter in magnitude and
fainter in radio flux. In spite of this difference, we see that the
overwhelming majority of the spectroscopic sample has $r > 30$,
i.e. the sample is made of radio-loud sources, regardless of the
nature of the radio emission.

Radio luminosities have then been derived according to the relation
${\cal P}\equiv{\rm P}_{1.4\rm GHz}=S_{1.4\rm GHz} D^2 (1+z)^{3+\alpha}$, 
and are expressed in [W Hz$^{-1}$ sr$^{-1}$] units. In the above formula, $D$
is the angular diameter distance and $\alpha$ is the spectral index of
the radio emission ($S(\nu)\propto \nu^{-\alpha}$).  Following
Magliocchetti et al. (2002), we assumed $\alpha=0.5$ for Seyfert 1
galaxies, $\alpha=0.75$ for early-type galaxies (with or without
emission lines due to AGN activity), $\alpha=0.7$ for Seyfert 2's and
$\alpha=0.35$ both for late-type galaxies and starbursts.  Fig.
\ref{fig:power_hist_type} shows the resulting distribution of radio
luminosities for the whole spectroscopic sample (lower panel) and for
the sub-class of radio-AGNs (top panel). The lack of objects at $\rm
log_{10} {\cal P} \simlt 21-21.5$ seen in the distribution of
AGN-fuelled sources is not seen in the whole spectroscopic sample
because of the population of low-luminosity star-forming galaxies
which are limited to radio powers $\rm log_{10} {\cal P} \simlt
22$ (see also Magliocchetti et al. 2002).  We also note that all the
sources in our sample have moderate luminosities (${\cal P}\simlt 10^{24}$), 
i.e. that all the
AGN-fuelled objects belong to the class of FRI galaxies (Fanaroff \&
Riley 1974). This is as expected, since the local radio luminosity
function of steep spectrum sources drops rapidly at ${\cal P}\simgt 10^{25}$ 
(Dunlop \& Peacock 1990),
which is the typical minimum luminosity of an FRII galaxy.
 
With the aim of studying the clustering properties of the population
of local radio sources, following Hawkins et al. (2003) and Madgwick
et al.  (2003) we have discarded from the original FIRST-2dFGRS sample
all the radio objects in 2dFGRS fields with $< 70$~\% completeness.
Furthermore, we have also only considered those sources with redshifts
$0.01<z<0.3$ and magnitudes $\bj\le 19.37$. This magnitude limit is
the brightest over the area of the 2dFGRS survey (Colless et al. 2001, 2003),
and so using this limit over the whole area removes the need to
correct for variations in the magnitude limit.  These cuts leave us
with a total of 761 objects (spectroscopic sample) out of which 536
belong to the population of AGN-fuelled sources (radio-AGN
sample). Their redshift distributions are illustrated by the
histograms in Fig. \ref{fig:N_z}.

\subsection{The Redshift-Space Correlation Function}

\begin{figure*}
\vspace{8cm}  
\includegraphics{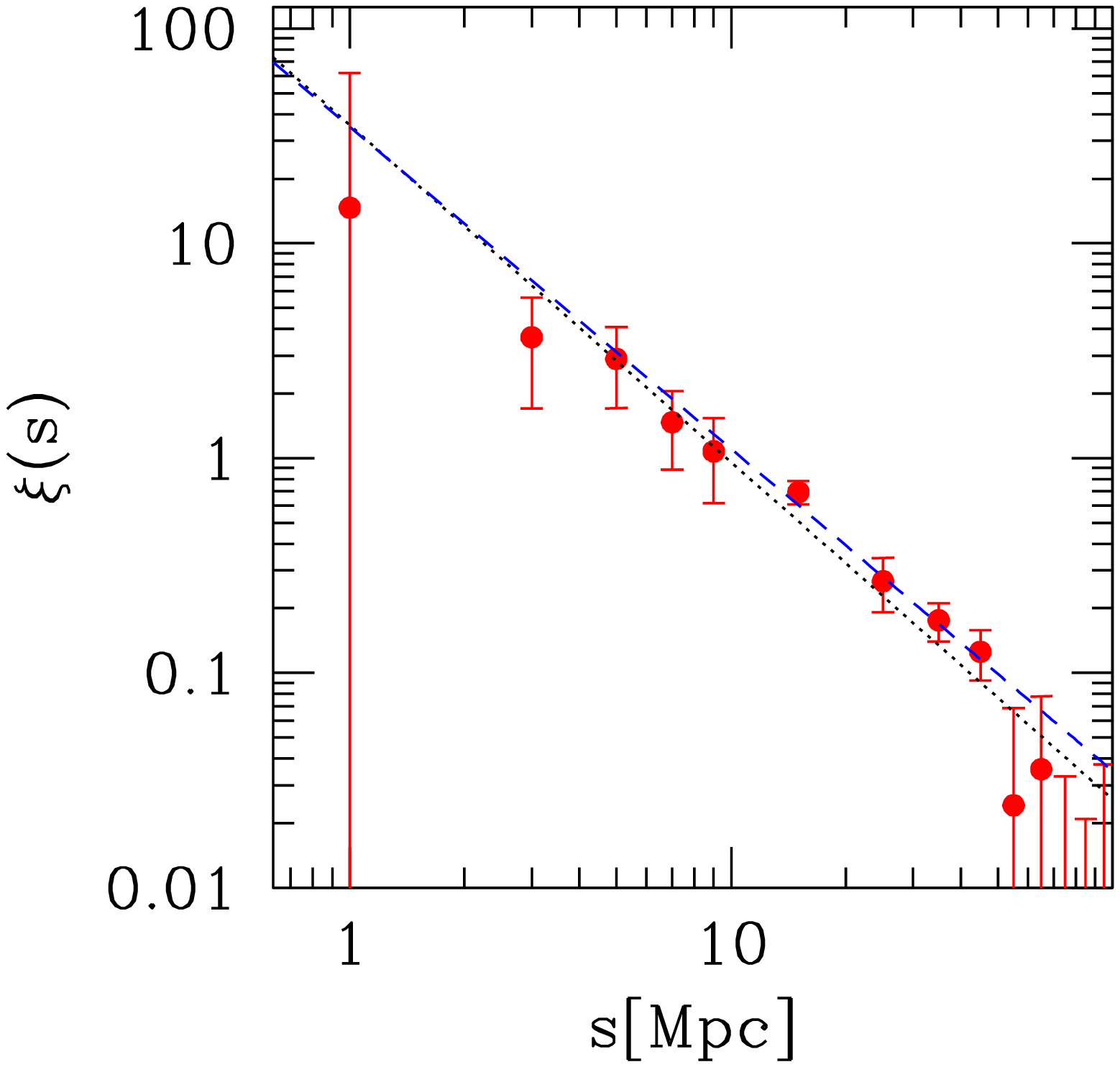}
\includegraphics{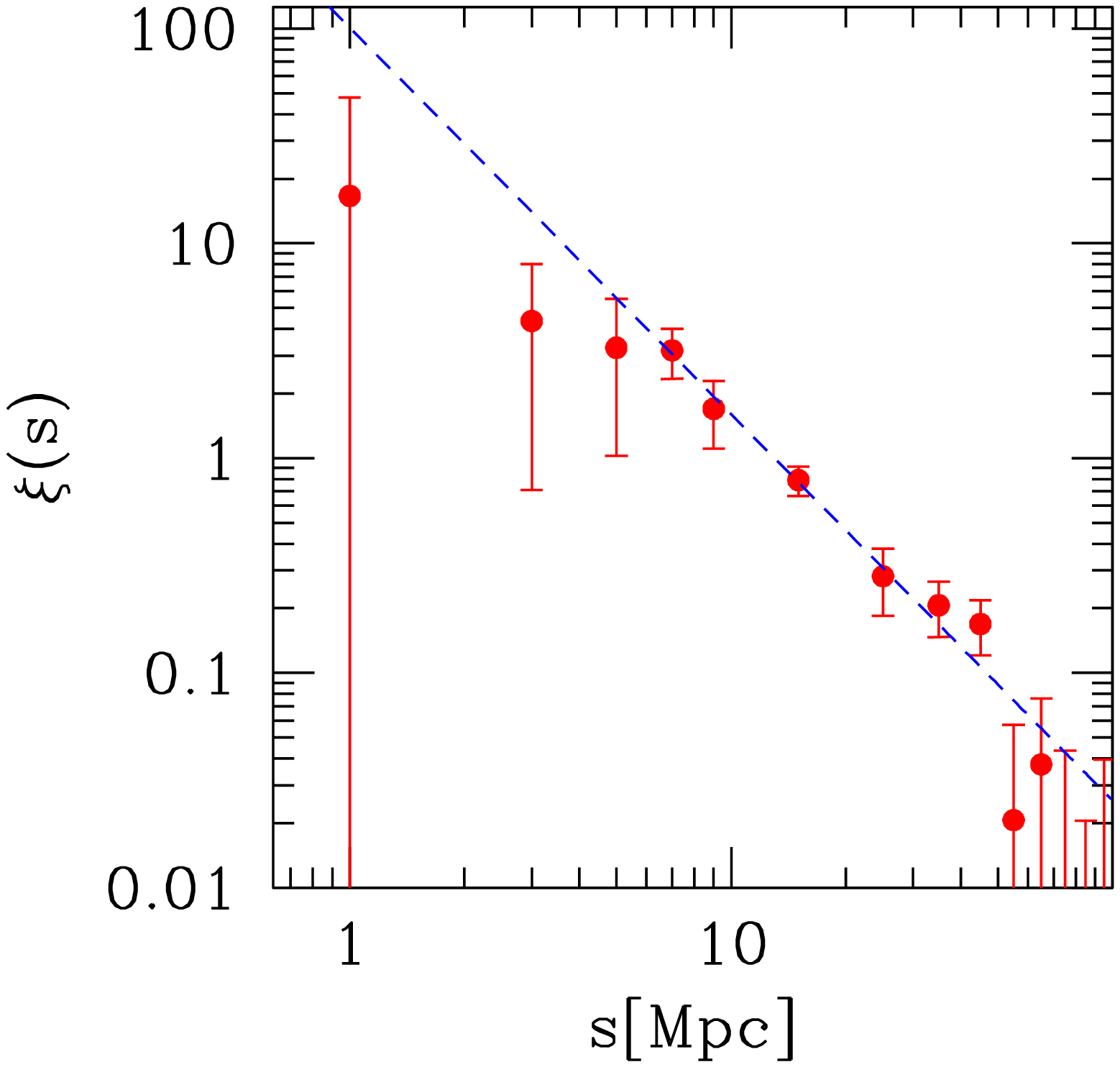}
\caption{Left-hand panel: redshift-space correlation function for the whole 
spectroscopic sample (761 sources). Error bars are obtained via jack-knife resampling 
of the dataset under consideration; given the large uncertainties associated with 
the smallest-scale measurement in the corresponding bin we only plot its upper limit. 
The dashed line 
represents the best fit to the data (derived only for those points 
included in the range $6 \simlt s/[{\rm Mpc}]\simlt 50$) corresponding to 
$\gamma_s=1.5$ and 
$s_0=10.7$~Mpc, while the dotted line illustrates the best-fit results found 
by Hawkins et al. (2003) for the entire population of 2dFGRS galaxies. 
Right-hand panel: redshift-space correlation function for the sub-sample of 
radio-AGNs (536 sources). Error bars are obtained as for the spectroscopic 
sample. The dashed line represents the best fit to the $6 \simlt s/[{\rm Mpc}]
\simlt 50$ data, corresponding to $\gamma_s=1.8$ and $s_0=13$~Mpc. 
\label{fig:xi_s}}
\end{figure*}

The standard way to quantify the clustering properties of a particular 
class of sources is by means of the two-point correlation function $\xi$ 
which measures the excess probability of finding a pair in the two volume 
elements $dV_1$ and $dV_2$ separated by a distance $r$. In practice, $\xi$ 
is obtained by comparing the actual source distribution with a catalogue of 
randomly distributed objects, subject to the same redshift and mask 
constraints as the real data.  We chose to use the estimator (Hamilton 1993)
\begin{eqnarray}
\xi(s) = 4\times \frac{DD\; RR}{(DR)^2} -1,
\label{eq:xiest}
\end{eqnarray}
where $DD$, $RR$ and $DR$ are the number of data-data, 
random-random and data-random pairs separated by a distance $s$ (note 
that throughout this section we will write the distance as $s$ instead 
of $r$ to stress the fact that we are working in redshift space). 

We generated random catalogues with ten times as many objects as the
real datasets, and modulated the angular distribution by both the
2dFGRS and FIRST coverage maps, so that the instrumental window
functions did not affect the measured clustering.  
The redshift distribution for these objects was then generated using
the selection function of the different datasets under consideration. Equivalent 
$w_i(z)=1/[1+4\pi\;n_R(z) J_3(s)]=1$ (with $n_R(z)$ number density of radio sources at 
redshift $z$ and $J_3\equiv \int_0^s \xi(s'){s'}^2 ds'$; 
see Hawkins 
et al. 2003 and Madgwick et al. 2003) weights were assigned to each object both in 
the real and random catalogues; this assumption was justified by the extremely low 
number density of radio sources ($n_R(z)^{max}\simeq 6\; 10^{-6}$~Mpc$^{-3}$) which makes 
the $4\pi\;n_R(z) J_3(s)$ contribution to $w_i$ a negligible one for any sensible value 
of $J_3$.

The above procedure was repeated both for the whole spectroscopic sample
and in the case of only radio-AGNs. The redshift distributions for the
two random catalogues corresponding to the different samples are shown
by the dashed lines in Fig. \ref{fig:N_z}.  The resulting
redshift-space correlation functions $\xi(s)$ are shown in Fig.
\ref{fig:xi_s}, where the left-hand panel is for the full
spectroscopic sample, and the right-hand panel for
radio-AGNs. The error bars on $\xi(s)$ were obtained by jack-knife
resampling each data-set 20 times.

Fig. \ref{fig:xi_s} suggests that a universal power-law is not a
good fit on all scales for either of the measured $\xi(s)$'s and that
it is particularly poor for the AGN-powered sources.  Although a
power-law fits well over the range $6 \simlt s/[{\rm Mpc}]\simlt 50$,
the redshift-space correlation function steepens at larger scales, and
flattens on smaller scales, particularly for the radio-AGN sample. We
estimate the true value of the redshift-space correlation length $s_0$
by fitting a localized power-law of the form
$\xi(s)=(s/s_0)^{-\gamma_s}$ only in the range $6 \simlt s/[{\rm
Mpc}]\simlt 50$.

The corresponding best values for the parameters $s_0$ and $\gamma_s$
can be obtained by a $\chi^2$ fit to $\xi$ as a function of $s$. This
analysis gives $s_0=10.7^{+0.8}_{-0.7}$~Mpc and $\gamma_s=1.5\pm 0.1$ 
for the whole FIRST-2dFGRS sample, while in the case of
AGN sources one gets $s_0=13\pm 0.9$~Mpc and
$\gamma_s=1.8^{+0.1}_{-0.2}$. These best-fits are shown by the dashed lines
in Fig. \ref{fig:xi_s}. The dotted line on the left-hand panel of
Fig. \ref{fig:xi_s} shows the redshift-space correlation function
for the entire population of 2dFGRS galaxies, derived by Hawkins et
al. (2003); the similarity between this measurement and the one for
local radio-sources is remarkable. This can be understood because the
FIRST-2dFGRS catalogue probes radio fluxes down to $S_{\rm 1.4
GHz}=1$~mJy, which results in a mixture of ``classical'' AGN-fuelled
sources, generally hosted by elliptical galaxies, and star-forming
objects such as spirals and irregulars (see Section 2). This mix of
early and late type galaxies makes the dataset look similar to a fair
(even though very sparse) sample of the whole population of 2dFGRS
galaxies.

The situation however looks very different if one considers only
AGN-fuelled sources. In this case the sample is much more strongly
correlated; the correlation length is higher, and also $\xi(s)$ has a
steeper slope. We will discuss the implications of these findings in
Section 3.3 when dealing with the real-space correlation function.

As a final step, we are interested in investigating whether there is
any dependence of the clustering signal on radio luminosity. We
divided the radio-AGN sample into two distinct subsets depending on
whether the sources have radio luminosity brighter or fainter than
${\rm log_{10}}{\cal P}=22.0$. There are
260 objects in the faint sample, and their average radio luminosity is
${\rm log_{10}}{\cal P}=22.9$. There are
276 objects in the bright sample, with average radio luminosity
${\rm log_{10}}{\cal P}=21.6$.

\begin{figure}
\vspace{8cm}  
\includegraphics{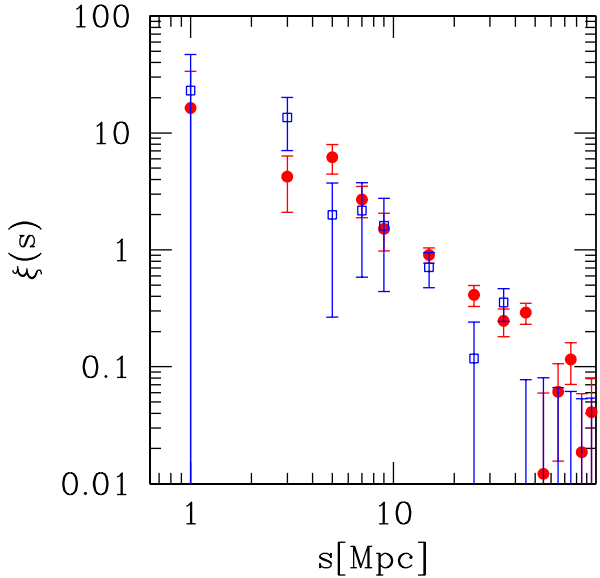}
\caption{Redshift-space correlation function for the two sub-samples
of radio-AGNs respectively brighter (open squares) and fainter (filled
circles) than ${\rm log_{10}}{\cal P} =22$.
Error bars are from Poisson noise estimates for the dataset under
consideration.
\label{fig:xi_type}}
\end{figure}

The redshift-space correlation functions for the two samples are plotted
in Fig. \ref{fig:xi_type}; the fainter sample is plotted as filled
circles, and the brighter sample as open squares. Even though the
small number of sources in the subsets lead to large uncertainties in
the measurements, it is clear that the two $\xi(s)$'s are entirely
compatible with each other. The best-fit parameters are $s_0 = 12\pm
3$~Mpc and $\gamma_s = 1.5\pm 0.15$ for brighter AGNs and $s_0 = 13 \pm 3$~Mpc
and $\gamma_s = 1.4 \pm 0.1$ for the fainter ones. This result shows
no evidence for  luminosity dependence in the clustering amplitude,
and shows that the amplitude of the correlation function at 10~Mpc
cannot differ by more much than a factor of 2 even though the two
samples differ by a factor of 20 in luminosity.  This is consistent
with the results of Peacock \& Nicholson (1991) who analysed an
all-sky sample of radio galaxies at $z \simlt 0.1$, and found no
evidence for any difference in clustering amplitude for samples $22.5
< {\rm log_{10}}{\cal P} < 23.5$ and $23.5 < {\rm log_{10}}{\cal P} < 24.5$.  
These results imply that the correlation amplitude of radio galaxies is
independent of their radio luminosity, at least in the power range
$10^{20}\simlt {\cal P} \simlt 10^{25}$.

\subsection{The Projected Correlation Function}

The previous Section has shown that the redshift-space correlation
function is extremely useful for characterizing the main properties of
the clustering of low-redshift radio sources.  However, $\xi(s)$
differs from the real-space correlation function because of peculiar
velocities that lead to redshift-space distortions (see e.g. Hawkins
et al. 2003 and Madgwick et al. 2003).  It follows that both the
amplitude and the slope of the real-space correlation function
$\xi(r)$ will differ from the values measured in Section 3.2.

The standard way to recover $\xi(r)$ is to first compute the
two-dimensional correlations as a function of separation parallel and
transverse to the line of sight, $\xi(\rp,\rt)$ and then integrate it
along the $\rp$ direction in order to obtain the projected correlation
function $\Xi$ as:
\begin{eqnarray}
\Xi(\rt)=2\;\int_0^\infty  \xi(\rp,\rt)\; d\rp
\label{eq:projxi}
\end{eqnarray}
We estimate $\xi(\rp,\rt)$ by means of equation (\ref{eq:xiest}) --
where this time pairs are counted in a grid of bins $\Delta \rp$,
$\Delta \rt$ -- and derive $\Xi$ by means of equation
(\ref{eq:projxi}). We set the upper limit for the integral
to $\rp^{\rm max}=60$~Mpc as the best compromise between a large
enough value to produce stable results and a small enough value to
avoid extra noise added to $\Xi$.

\begin{figure*}
\vspace{8cm}  
\includegraphics{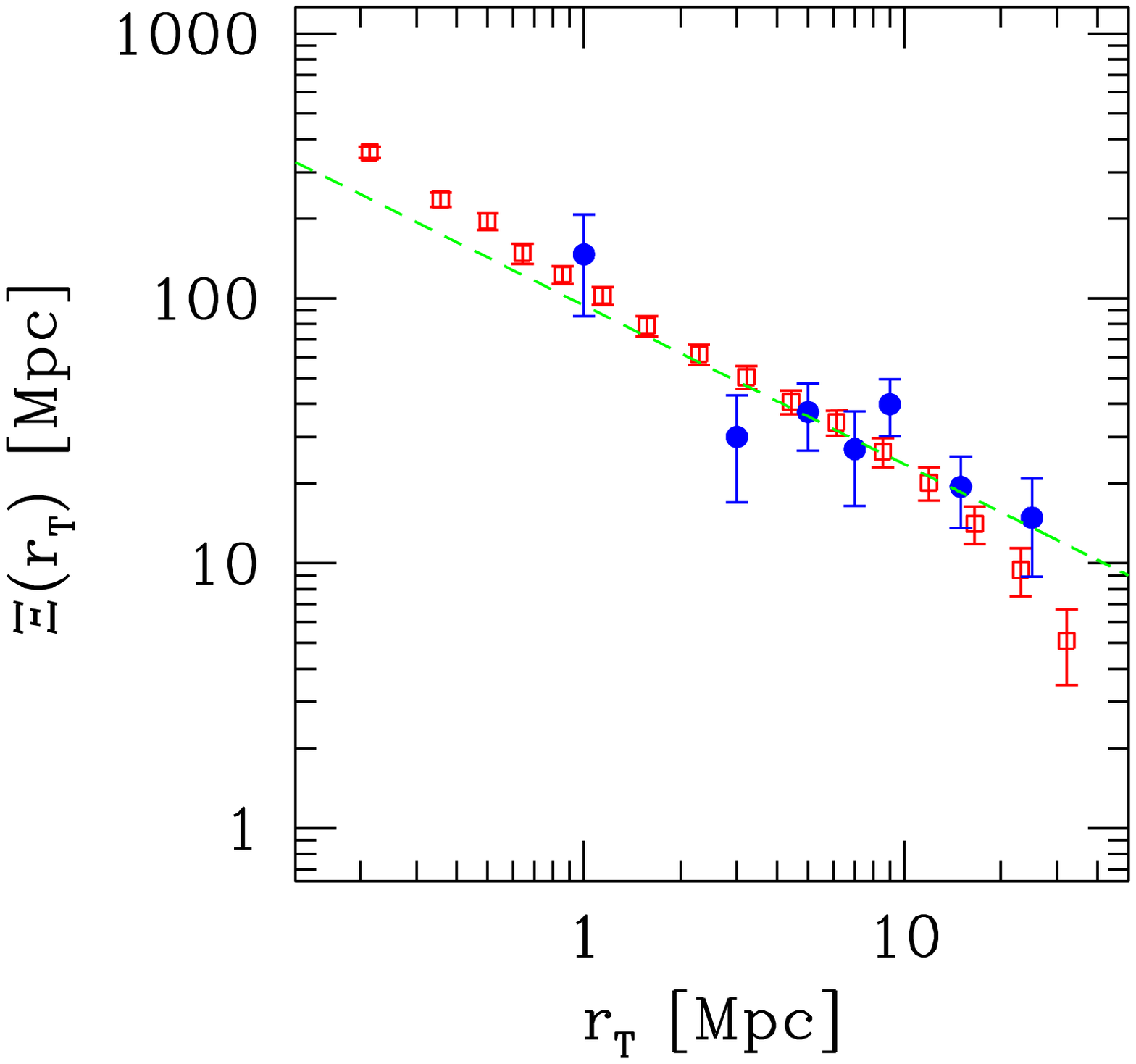}
\includegraphics{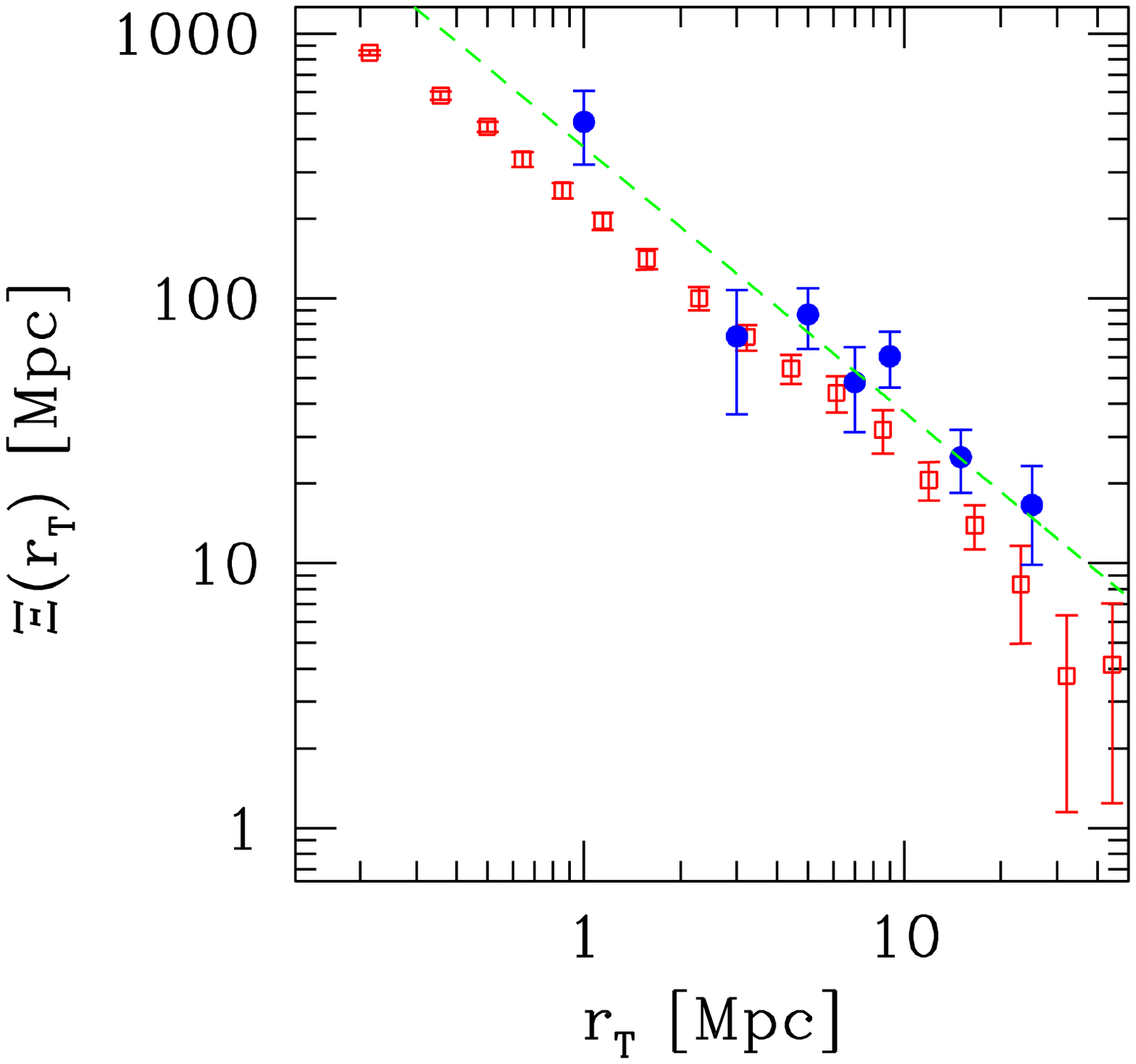}
\caption{Left-hand panel: projected correlation function for the whole 
spectroscopic sample (761 sources). Error bars are obtained as 
jack-knife resampling of the dataset under consideration. The dashed line 
represents the best fit to the data corresponding to $\gamma=1.6$ and 
$r_0=6.7$~Mpc, while open squares illustrate the results obtained by Hawkins 
et al. (2003) for the entire population of 2dFGRS galaxies. 
Right-hand panel: projected correlation function for the sub-sample of 
radio-AGNs (536 sources). Error bars are obtained as for the spectroscopic 
sample. The dashed line represents the best fit to the data corresponding to 
$\gamma=2.0$ and 
$r_0=10.9$~Mpc, while open squares illustrate the results obtained by Madgwick 
et al. (2003) for the population of early-type 2dFGRS galaxies. 
\label{fig:proj_xi}}
\end{figure*}

As in the previous subsection, $\Xi$ has been evaluated both using the
whole spectroscopic sample and the sample of objects showing
signatures of AGN in their spectra. Once again, the errors were
calculated from 20 jack-knife resamplings of the datasets under
consideration. 

The resulting measurements on scales 1-30 Mpc are
shown by the solid points in the two panels of Fig.
(\ref{fig:proj_xi}): the left panel shows the results for whole
spectroscopic sample, and right panel show the results for AGN sample.
The open squares in Fig. (\ref{fig:proj_xi}) show the projected
correlation functions from corresponding optical 2dFGRS samples: in
the left panel they show the results for the entire population of 2dF
galaxies (Hawkins et al. 2003); in the right panel they show the
results for early-type 2dFGRS galaxies (Madgwick et al. 2003).  From
these comparisons we see that the clustering properties of the
whole spectroscopic sample are remarkably similar to those of
``normal'' galaxies, but the projected correlation function of radio
AGNs has an amplitude that is roughly twice as large as that of local
ellipticals.

From the measurements presented in Fig. \ref{fig:proj_xi} it is then
possible to estimate the real-space correlation function $\xi(r)$ via
(Davis \& Peebles, 1983)
\begin{eqnarray}
\Xi(\rt)=2\int_{\rt}^{\infty}\xi(r)\frac{r\; dr}{(r^2-\rt^2)^{1/2}}
\label{eq:xibar}
\end{eqnarray}
which, if we assume the power-law form $\xi(r)=(r/r_0)^{-\gamma}$, 
can be integrated analytically leading to 
\begin{eqnarray}
 \Xi(\rt)=r_0^\gamma \rt^{1-\gamma} H_\gamma,
\end{eqnarray}
where $H_\gamma=\Gamma(\frac{1}{2})\Gamma(\frac{\gamma-1}{2})/\Gamma
(\frac{\gamma}{2})$.

A least-square fit to the data for the whole spectroscopic sample
gives $r_0=6.7^{+0.9}_{-1.1}$~Mpc, $\gamma=1.6 \pm 0.1$, surprisingly
close to the results of Hawkins et al. 2003.  For the AGN population
we find $r_0=10.9^{+1.0}_{-1.2}$~Mpc, $\gamma=2.0\pm 0.1$, which
is similar to that found by Norberg et al. (2002) for the brightest
early-type galaxies in the 2dFGRS.
A summary of the results derived here and in Section 3.2 can be found
in Table \ref{table1}.

We note that a measured correlation length $r_0\simeq 11$~Mpc is in very good agreement 
with previous estimates derived from deprojection of the angular two-point correlation function 
$w(\theta)$. These include the results of Cress et al. (1996), Loan et al. (1997), Magliocchetti 
et al. (1998) and Magliocchetti \& Maddox (2002). However, more recent works (e.g. Blake \& Wall 
2002; Overzier et al. 2003) seem to find lower values for $r_0$, of the order of 5-6~$h^{-1}$
~Mpc. Given that the functional form for the redshift distribution of radio sources at the mJy 
level adopted by the various authors is the same, i.e. the one proposed by Dunlop \& Peacock 
(1990), we ascribe the differences found by Blake \& Wall (2002) and Overzier et al. (2003) as 
caused by not properly dealing with the issue of multi-component sources. Ignoring the effect of 
multiple-sources on the observed clustering properties of radio objects in fact leads (amongst 
others) to an underestimate of the slope of $w(\theta)$, $1-\gamma$, which in turn yelds
smaller inferred values for the corelation length $r_0$.

\begin{table}
\begin{center}
\caption{Best-fit parameters to $\xi$. In the case of $\xi(s)=(s/s_0)^
{-\gamma_s}$ the fit only uses points with $6\simlt s/[{\rm Mpc}] \simlt 50$
~Mpc, while for $r_0$ 
and $\gamma$ all the points between $\rt=1$ and $\rt=30$ Mpc have been 
considered.
\label{table1}}
\begin{tabular}{lll}
\hline
       & Whole Sample & Radio-AGN\\
\hline
\hline
$s_0$ [Mpc]&$10.7^{+0.8}_{-0.7}$ & $13\pm 0.9$\\
$\gamma_s$ &$1.5\pm 0.1$ & $1.8^{+0.1}_{-0.2}$\\
\hline
$r_0$ [Mpc] &$6.7^{+0.9}_{-1.1}$&$10.9^{+1.0}_{-1.2}$\\
$\gamma$ &$1.6\pm 0.1$ &$2.0\pm 0.1$\\
\end{tabular}
\end{center}
\end{table}

\section{Constraints on the Halo and Black Hole Mass}

Since the $\Xi$ measurements presented in the right-hand panel of
Fig. \ref{fig:proj_xi} have been obtained for a homogeneous sample
of AGN-fuelled sources, we can easily predict the clustering for a
range of models to be directly compared with the data.

Under the assumption that each dark matter halo can only
host one radio-AGN (but see later for a more general discussion), the predicted spatial 
correlation function $\xi$
(on a scale $r$ and at a redshift $z$) for this class of sources can
in fact be written as:
\begin{eqnarray}
\xi(r,z)=\xi_{\rm dm}(r,z)\;b^2_{\rm eff}({\rm M_{\rm min}},z).
\label{eq:xith}
\end{eqnarray}
This expression holds on scales $\simgt 1$~Mpc where the halo-halo
exclusion effects are negligible (for the general case see
Magliocchetti \& Porciani 2003).

The mass-mass correlation function $\xi_{\rm dm}(r,z)$ in Equation
(\ref{eq:xith}) is fully specified for a given cosmological model and
a chosen normalization $\sigma_8$. The precise form of $\xi_{\rm
dm}(r,z)$ can be analytically derived following the approach of
Peacock and Dodds (1996).  The effective bias $b_{\rm eff}({\rm M_{\rm
min}},z)$, which determines the way radio sources trace the mass
distribution, is obtained via:
\begin{eqnarray}
b_{\rm eff}({\rm M_{\rm min}},z)=\left[\int_{{\rm M_{\rm min}}}^\infty b(m,z)
\;n(m,z)\; dm\right]/ \nonumber\\
\left[\int_{{\rm M_{\rm min}}}^\infty n(m,z)\;dm\right],
\end{eqnarray}
where $b(m,z)$ and $n(m,z)$ respectively are the linear bias factor and 
the halo mass function of individual halos of mass $m$ at a redshift $z$ and 
-- in our specific case -- ${\rm M_{\rm min}}$ is the minimum mass of a halo able to 
host a radio-AGN. 

In order to compare models with data we have then written both
$b(m,z)$ and $n(m,z)$ according to the Sheth \& Tormen (1999)
prescriptions and derived the theoretical projected $\Xi$ by
integrating the spatial correlation function (\ref{eq:xith}) along the
direction parallel to the line of sight $\rp$ via equation
(\ref{eq:xibar}), where all the relevant quantities are evaluated at
the median redshift of the 2dFGRS, $\langle z \rangle\simeq 0.1$.

\begin{figure}
\vspace{8cm}  
\includegraphics{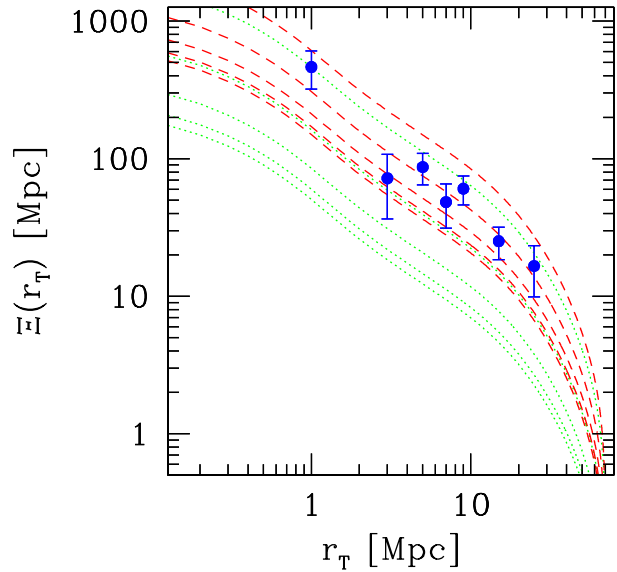}
\caption{Projected correlation function of radio-AGNs. Measurements
are those obtained in Section 3.3, while the dotted lines show the
predictions derived from equation (\ref{eq:xith}) for minimum halo
masses ranging from 10$^{10}$ (bottom curve) to 10$^{14}$ (top curve)
M$_{\sun}$. Dashed lines are obtained as in the former case but for a mean 
halo occupation number $\langle {\rm N_{radio}}\rangle \propto m$ (see text for details).
\label{fig:xi_th}}
\end{figure}

The dotted lines in Fig. (\ref{fig:xi_th}) show the predictions obtained from equation
(\ref{eq:xith}) for different values of the minimum halo mass
necessary to host a radio-AGN as compared to the measurements derived
in Section 3.3. The chosen values for the minimum mass range from
${\rm M_{\rm min}}=10^{10}$~M$_{\sun}$ (bottom dotted curve) to
10$^{14}$~M$_{\sun}$ (top dotted curve). Despite the large error bars, it is
clear that none of the predicted $\Xi$'s for  minimum halo
masses $\simlt 10^{13}$~ M$_{\sun}$ can provide a reasonable fit to
the data, since they all fall below the observed values. Indeed, the
best fit is obtained for ${\rm log_{10}}[{\rm M_{\rm min}}/{\rm
M}_{\sun}]=13.7^{+0.2}_{-0.3}$, corresponding to an effective bias $b_{\rm
eff}=1.85$. We note that very similar values (i.e. ${\rm log_{10}}[{\rm M_{\rm min}}/{\rm
M}_{\sun}]=13.5^{+0.2}_{-0.3}$ and ${\rm log_{10}}[{\rm M_{\rm min}}/{\rm
M}_{\sun}]=13.0^{+0.3}_{-0.5}$) are also found if one releases the 
assumption of one radio galaxy per halo and considers more general distributions 
of the kind $\langle {\rm N_{radio}}\rangle \propto m^{0.5}$ (see e.g. Peacock 2003) or a 
more extreme $\langle {\rm N_{radio}}\rangle \propto m$ 
(with $\langle \rm N_{radio}\rangle$ average number of radio galaxies per halo 
of mass $m$) even though, as expected, the difference between models derived for 
different values of $\rm M_{min}$ decreases as one increases the weight given to 
larger-mass haloes (see Fig. \ref{fig:xi_th}).

The implications of the above results are quite intriguing since they
suggest, independent of the chosen halo occupation form, that radio-AGNs reside in 
relatively massive haloes, spanning
a mass range from rich groups to super-massive clusters. Smaller halos
seem to be inhibited from hosting a radio-emitting AGN. This
conclusion is also supported by the small-scale behaviour of the
redshift-space correlation function of radio-AGNs as measured in
Section 3.2. Indeed, both the lack of pairs with distance $\simlt
1$~Mpc and the flattening of $\xi(s)$ on scales $\simlt 5$~Mpc have a
natural explanation if one assumes AGN-powered sources to be confined
in massive halos (even if one cannot exclude this latter finding as due to the fingers 
of God effect, see Hawkins et al. 2003). 
In the case where only one source per halo is assumed,
the halo-halo spatial exclusion ensures that no
pair of radio-AGNs can be found with a separation smaller than the sum
of the virial radii of the halos that host them.  For a typical halo
mass of $10^{13.4}$M$_{\sun}$ the minimum separation for the adopted
cosmology turns out to be $\sim$~1.6 Mpc.  Allowing for the
uncertainties in our determination of ${\rm M_{\rm min}}$, this implies that
there should be a deficit of pairs on scales $\sim 1-5$~Mpc, consistent
with the observed flattening of $\xi(s)$ on small scales.

Thus our average estimate of the minimum halo mass for a radio-AGN is ${\rm
log_{10}}[{\rm M_{\rm min}}/{\rm M}_{\sun}]\simeq 13.4$), which is about a
factor 10 larger than that associated with ``normal'' early-type
galaxies, ${\rm log_{10}}[{\rm M_{\rm min}}/{\rm M}_{\sun}]\simeq 12.5$,
(Magliocchetti \& Porciani 2003). This is also about a factor 10
larger than that obtained for the population of radio-quiet quasars in
the 2dF QSO dataset, ${\rm log_{10}}[{\rm M_{\rm min}}/{\rm M}_{\sun}]\simeq
12.1-12.6$ (Grazian et al. 2003).  All three of the above
determinations are based on samples drawn from the same parent
catalogue (APM survey), and so selection effects are not expected to
bias the comparison between the different results.\\ Taking these
numbers at face value suggests that radio-AGNs and radio-quiet quasars
are two distinct populations, with more massive dark matter halos
required for an AGN to trigger radio activity.
 
On the other hand, the results presented in Section 3.2 (see Fig.
\ref{fig:xi_type}) together with those obtained by Peacock \&
Nicholson (1991) for more luminous radio galaxies show that radio-AGNs
with very different radio luminosities have very similar correlation
amplitude. If one thinks in terms of the model expressed in equation
(\ref{eq:xith}), there appears to be little or no correlation between
radio luminosity and minimum halo mass. A simple comparison of Figs
6 and 8 illustrates that the correlation function estimates obtained
for low-luminosity and moderately bright-luminosity radio-AGNs both fall in the
same $\rm{10^{13}\simlt M_{\rm min}/M_{\sun}\simlt 10^{14}}$ range as found
for the whole sample).

If we then combine the above information, the
data seem to suggest that {\it there is a threshold halo mass
required to produce significant radio emission from an AGN but -- once
the radio activity is triggered -- there seems to be no connection
between radio luminosity of the sources and dark matter content of
those halos hosting them}. Note that this statement does not
necessarily imply the minimum halo mass to be the only factor playing
a relevant role in the onset of radio emission from an AGN. Also the
threshold may not be a sharp mass threshold, but could be as a smooth
transition between the radio-quiet and radio-loud regimes for halos
near $\rm M_{\rm min}$.

These results on the minimum halo mass can also be viewed in terms of
a minimum black hole mass. Ferrarese (2002) gives a simple
prescription to convert the dark matter mass $\rm M_{DM}$ of a halo
hosting a black hole into black hole mass $\rm M_{BH}$: 
\begin{eqnarray}
\frac{\rm M_{BH}}{\rm 10^8 M_{\sun} }\simeq 0.10 \left(\frac{\rm M_{DM}}
{10^{12}\rm M_{\sun}}\right)^{1.65}.
\label{eq:mbh}
\end{eqnarray}
Thus, a minimum halo mass ${\rm M_{\rm min}}\sim 10^{13.4} \rm M{\sun}$
corresponds to $\rm M^{\rm BH}_{\rm min}\sim 10^9 \rm M_{\sun}$, which
would represent the minimum black hole mass required to onset
radio-AGN activity -- at least for relatively faint/FRI sources such
as the overwhelming majority of those we have considered in our
analysis (see Section 3.1).  It is interesting to note that this
estimate is in good agreement with both the results of Marchesini,
Ferrarese \& Celotti (2003) and also those of Dunlop et al.  (2003),
despite the entirely different methods used in our analysis. Also our
minimum black hole mass is noticeably larger than that obtained for
radio-quiet quasars; from an analysis of the 2dF QSO dataset in the
same redshift range spanned by the sources considered in this work,
Corbett et al. (2003) find $6.5\simlt {\rm log_{10}}[{\rm M_{\rm BH}}/{\rm
M_{\sun}}]\simlt 8.3$.  By making use of equation (\ref{eq:mbh}), we
can then conclude that -- at least for moderately faint/FRI radio
sources -- {\it there seems to be a minimum black hole mass of about
$10^9 \rm M_{\sun}$ required to produce significant radio emission
from an AGN; however, once the radio emission is produced, there is
little or no correlation between radio luminosity and black hole
mass}.  This is in agreement with other analyses of radio-loud quasars
by Dunlop et al. (2003) and Cirasuolo et al. (2003).

We note that, even though the connection between black hole mass and radio power has been 
obtained in our analysis as a consequence of the connection between radio power and halo mass, 
this first relationship is supposed to be more fundamental and physically based as it is the 
black hole which actually powers the AGN. Any correlation between radio power and halo mass is 
then expected to arise as a secondary effect, since black hole mass is observed to scale with 
bulge luminosity and this is turn is related  -- in the case of elliptical galaxies -- to halo 
mass (see e.g. Magorrian et al. 1998; Kormendy \& Gebhardt 2001; McLure \& Dunlop 2002; 
Archibald et al. 2002; Granato et al. 2004)

\section{conclusions}

This paper has presented the clustering properties of local, $S_{1.4
\rm GHz}\ge 1$~mJy radio sources by making use of a sample of 820
objects drawn from the joint use of the FIRST survey and the
2dFGRS. To this aim, we have introduced 271 new $\bj \le 19.45$
spectroscopic counterparts for FIRST radio sources to be added to
those already obtained by Magliocchetti et al. (2002). These objects
can be divided in two broad sub-classes: (1) star-forming objects (66
sources, which include both late-type and star-burst galaxies), which
owe their radio activity to processes connected to intense star
formation and (2) radio-AGNs (202 sources), where the radio signal
stems from accretion processes onto a central black hole. In both
cases, the redshift range range spanned extends up to $z\simeq 0.3$
(with star-forming objects being relatively more local than
AGN-fuelled sources) and radio luminosities cover the interval
$10^{20} \simlt {\cal P} \simlt 10^{24}$,
which identifies the radio-AGNs included in the sample as FRI sources.

The redshift-space correlation function $\xi(s)$ and the projected
correlation function $\Xi(\rt)$ have been calculated for both the
total FIRST-2dFGRS sample and for the sub-class of radio-AGNs. The 
results for the two populations are quite different.

In the case of all FIRST-2dFGRS sources, both the correlation functions
are found to be entirely consistent with the estimates obtained by
Hawkins et al. (2003) for the whole sample of 2dFGRS galaxies. From
measurements of the redshift-space correlation function
$\xi(s)=(s/s_0)^{- \gamma_s}$ we derive a redshift-space clustering
length $s_0=10.7^{+0.8}_{-0.7}$~Mpc and a slope $\gamma_s=
1.5\pm 0.1$, while from the projected correlation function
$\Xi(\rt)$ we estimate the parameters of the real-space correlation
function $\xi(r)=(r/r_0)^ {-\gamma}$ to be $r_0=
6.7^{+0.9}_{-1.1}$~Mpc and $\gamma= 1.6\pm 0.1$.

Sources that show signatures of AGN activity in their spectra are
found to be very strongly clustered, with a value for the
redshift-space clustering length, $s_0=13.0\pm 0.9$ (in
excellent agreement with the findings of Peacock \& Nicholson
1991). Measurements of the real-space correlation function lead to
$r_0\simeq 11$~Mpc and $\gamma\simeq 2$, slightly steeper and higher
amplitude than optically selected galaxies, but very similar to bright
early-type galaxies.

We also find no significant differences in the clustering properties
of faint (${\cal P}\le 10^{22}$) compared to brighter (${\cal P} > 10^{22}$)
AGN-fuelled radio sources.

Comparisons with physically-motivated models for the clustering
properties of classes of galaxies show that AGN-fuelled sources have
to reside in dark matter halos more massive than $\rm
log_{10}[M_{min}/M_{\sun}]\simeq 13.4$, higher than the figure recently
measured for radio-quiet QSOs (see e.g. Grazian et al. 2003). Under
certain assumptions, this value can be converted into a minimum black
hole mass associated with radio-loud, AGN-fuelled objects of $\rm
M_{BH}^{min}\sim 10^9 M_{\sun}$, again larger than current estimates
for the typical black hole mass associated with local 2dF (radio-quiet)
quasars (Corbett et al. 2003).

The above results then suggest -- at least for moderately faint/FRI
radio sources such as those included in our sample -- the existence of
a threshold halo/black hole mass associated with the onset of
significant radio activity such as that of radio-loud AGNs.  We stress
that such threshold is not necessarily a sharp one, as it could as well
identify an allowed range (in the high-mass regime) for the transition
between radio-quiet and radio-loud regimes.  Once the activity is
triggered there seems to be no evidence for a connection between
halo/black hole mass and radio luminosity.

\end{document}